# Fostering Student Enrollment in Basic Sciences: the Case of Southern Tuscany

Vera MONTALBANO
Department of Physics, University of Siena
Siena, Italy

## ABSTRACT

In recent decades it has been detected in Italy a decrease in enrollment in basic sciences, i.e. Mathematics, Physics and Chemistry. The increase in specific orientation in Chemistry, Physics, Mathematics and Materials Science, is strategically crucial to achieve the goal of maintaining and increasing the number of motivated and capable students who enroll in these and other courses scientific degree.
In 2005 the government launched the Scientific Degree Project in which many Italian Universities were involved in implementing activities focused to enhance the interest of high school students towards sciences and scientific degrees. With the purpose of increasing scientific vocations, workshops were organized in high schools and teachers were involved in planning and implementation of laboratories, conferences for scientific outreach, thematic exhibitions, guided tours of research laboratories, summer's schools for students and courses for teachers were realized for developing a cultural enhancement in teaching basic sciences. Initially, the project had been funded for four years, then was refinanced in 2010 and became the National Plan for Science Degree, with the aim of stabilizing the most significant activities and promoting them throughout the country. Particularly significant is the case of activities organized by the Department of Physics of the University of Siena for students and teachers in Southern Tuscany. The methods used in cultural enhancement of teachers and activities designed to support schools with limited laboratory facilities, together with stimulating activities for the more motivated and talented students are allowed to take root for some good practices in physics teaching and orientation to scientific degrees. The project was born in Siena only for Physics in 2005, adding the project for Chemistry in 2007 and in 2009 also the Department of Mathematical Sciences launches its project. Beyond describing the main activities for orientation to Physics, activities done in partnership with chemists, biologists and geologists are reported, as well as an activity in which the Departments of Mathematical Sciences and Physics are both involved in looking for introducing new interdisciplinary methodologies to increase students' understanding in high school of some selected topics in which both Mathematics and Physics give a contribution in the construction of important and mutually reinforcing basic concepts of the two disciplines.

## 1. INTRODUCTION

The enrollment in Physics, Mathematics and Chemistry has declined dramatically almost everywhere in the world [1-5]. The interest of students in learning these disciplines decreases and consequently the results achieved in schools and universities are disappointing. On the other side, the goal of maintaining and increasing the number of motivated and talented students who enroll in courses scientific degrees is crucial in order to achieve a knowledge-based economy more competitive and dynamic capable of sustainable economic growth with more and better jobs and greater social cohesion.

Italy is active since 2005 with a large and detailed plan to remedy this problem funded by the Ministry of Education and Scientific Research [6]. The main purpose is to provide opportunity for students in the last years of high school to learn about issues, problems and processes characteristic of scientific knowledge, also in relation to the areas of labor and the professions, in order to identify specific interests and requirements, making informed and suitable choices in connection to their personal project. Another key objective is to improve the disciplinary and interdisciplinary knowledge of teachers and their ability to interest and motivate students in learning science, and to support them in pre-university orientation process.

In paragraph 2, the enrollment trend in Italy is presented. Actions for inverting this trend are described with a special care to strategy and adopted methodologies. In the next paragraph, the case of southern Tuscany is discussed in detail. Many designed and realized activities are reported. In particular, a summer's school for students is described. Its purpose is to orient students towards physics and to improve teacher practice.

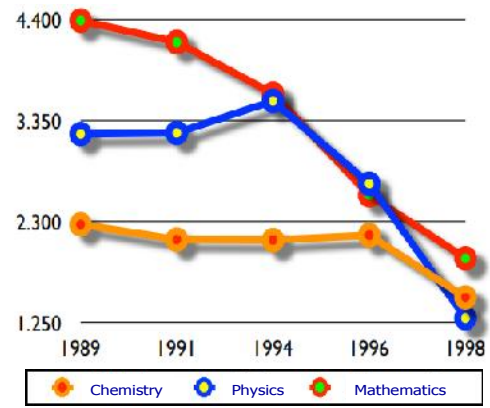

*Figure 1: Enrollment in Italian universities before actions were performed for inverting the negative trend [7]*

Finally, for teacher professional development, an activity for teachers of physics and mathematics on modeling is presented. A brief account of the effect of these actions on the enrollment trend concludes this article.

## 2. ACTIONS FOR SCIENCE DEGREE IN ITALY

In recent decades it has been detected almost everywhere in the world a consistent decrease in enrollment in basic sciences, i.e.

Mathematics, Physics and Chemistry. The main consequence is a decreasing of graduates in science disciplines.

The situation in Italy was dramatic: Figure 1 shows that enrollments in basic sciences are more than halved in few years. Since there is a clear association between economic performance and the numbers of engineers and scientists produced by a society, the Ministry of Education and Scientific Research promoted a wide project in order to reverse this trend.

Starting at the end of 2005, the project was named Scientific Degree Project ( Progetto nazionale per le Lauree Scientifiche, i.e. PLS) and was financed for four years. During this period and at the end of the project, a large monitoring of all activities was realized in order to identify what actions were more effective and incisive. In 2009 it was launched the National Plan for Science Degree (the same acronym remains: PLS) where some of the most effective methodological aspects were emphasized in new guidelines.

|  | **Scientific Degree Project 2006 - 2009** | **National Plan for Science Degree 2010 - 2012** |
|---|---|---|
| funds from Ministry | 8,5 M€ | 4,8 M€ |
| funds from Universities | 2,7 M€ | 0,5 M€ |
| Universities | 33 | 42 |

*Table 1: Financial support in Italy from 2006 to 2012 and universities involved in contrasting decreasing of scientific vocations.*

The source of main financial support for both actions is showed in Table 1, together with the participating universities.

**Scientific Degree Project 2006 - 2009**
The project stems from a collaboration of the Ministry of Education and Scientific Research, the National Conference of Deans of Science and Technology and Confindustria, the main organisation representing Italian manufacturing and services companies, and was designed in 2004 with the initial motivation to increase the number of students on degree courses in Chemistry, Physics, Mathematics and Science of materials.

The project focused on three main objectives [6]:
 - improving knowledge and awareness of science degree in secondary school, offering students in the last three years of school to participate in stimulating and engaging laboratory activities curricular and extra curricular;
 - starting a process of professional development of science teachers in service in the Secondary School from joint work between the School and University for the design, implementation, documentation and evaluation of the laboratories mentioned above;
 - promote alignment and optimization of training from the University and the University School for the working world, strengthening and stimulating activities and training workshops at universities, research institutions, public and private companies engaged in Research and Development.

The action for student integrated witn training of teachers was made through more than 100 sub-projects under the responsibility of local referents, located in 33 universities, became 38 in the last period, spread all over the country and organized into four areas of national projects [7], as shown in Figure 2.

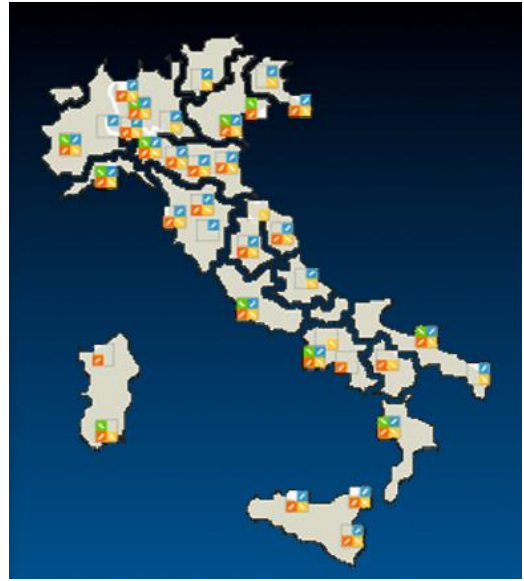

*Figure 2: Universities in the national project. For each university the activated areas of the project are shown: materials science, physics, mathematics and chemistry (clockwise)*

The main novelty compared to previous actions consists in using the same organizational model coordinated at national level, introducing a new model of collaboration between universities and secondary schools and the involvement of new actors in the activity of connecting university – school.

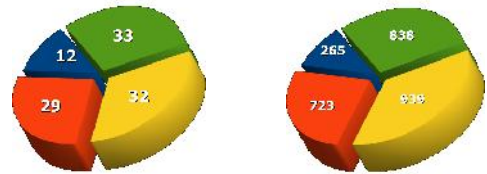

*Figure 3: Universities and schools involved in disciplinary projects, starting from smaller, science of materials, chemistry, physics and mathematics (clockwise)*

Educational institutions involved in disciplinary projects [7]. are shown at the end of fist biennial period in Figure 3.

**National Plan for Science Degree 2010 - 2012**
The plan maintains the same purposes of increasing the enrollment in science degrees to which is added the necessity to revise the content and methods of teaching and learning of science in all grades of school, taking into account the new national guidelines for first and second cycle contained in the recent Italian reform of the educational system.

 **Strategy and methodologies:** In order to achieve the above purposes, the Plan pursuits fundamental ideas that have proved effective in trials 2005-2009 [6]:
 - orientation does not conceive how a teaching path given to student, but as an action that the student is doing, from meaningful activities that allow to compare problems, issues and ideas of science;
 - designing the training of teachers in service by involving teachers in solving concrete problems, developing design and

implementation of educational activities and through comparison with peers and experts;
 - pursuing and achieving at the same time the student orientation and training of teachers through the planning and joint implementation by school teachers and university laboratories for students, thus developing also relations between the school system and the University;

Furthermore, a new idea adds: connecting consciously the activities of the Plan with the innovation of curricula and teaching methods adopted in schools, and other contents and methods of teacher training (initial and in-service), for the first and second cycle.

In other words, the main road consists in considering laboratory as a method, not as a place, students must become the main character of learning and joint planning by teachers and university is a mandatory step.

More attention to laboratory is required and different type of laboratory PLS can be proposed:
  - laboratories which approach the discipline and develop vocations,
  - self-assessment laboratories for improving the standard required by graduate courses,
  - deepening laboratory for motivated and talented students.

### 3. ACTIONS IN SOUTHERN TUSCANY

During the last decade, the Department of Physics is active in orienting students towards Physics and in teachers training in Physics and Mathematics. Our activities are directed to students and teachers living in the provinces of Arezzo, Grosseto and Siena, showed in Figure 4.

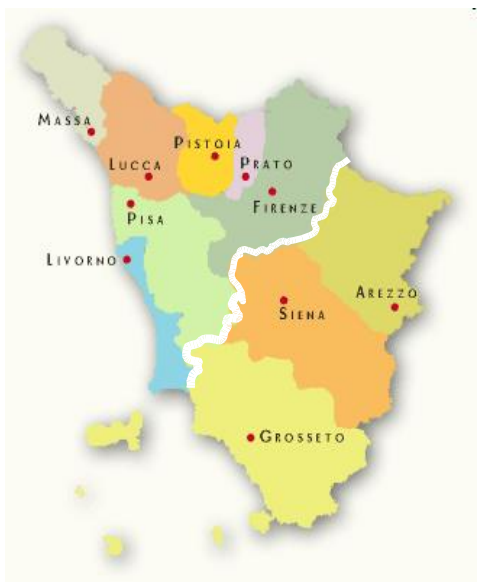

*Figure 4: the area in Tuscany where Department of Physics performs its activities of orienting and teacher professional training and development.*

**Early activities**
Since 2000, the Department of Physics is engaged in many activities aimed at students and teachers of high schools, such as initial teacher training in the Advanced School for Teaching in Secondary School of Tuscany and various activities for orienting students towards Physics. Furthermore, an advanced courses for teachers of Physics and Mathematics was held. In this course, teachers were engaged in designing learning paths with a particular attention to relationship between physics and mathematics. Physicists (E. Mariotti, V. Millucci, V. Montalbano, S. Veronesi) and expert teachers (R. Benedetti, A. Porri, E. Papi) worked together at that time in order to improve teaching strategy and methodologies in these disciplines..

**Scientific Degree Project**
The project started in Siena only for Physics in late 2005 with 11 school involved, while the project for Chemistry began in 2007. Physics project was named *Trying and trying again* for remembering to everyone that main activities were focused on laboratories directly performed by students or teachers in the case of professional development. Two main actions, professional development for teachers and orientation to physics for students, were implemented and in the following few examples of both are given.

**Advanced course for professional development:** We continued a previous experience by designing an annual course for teachers titled *Learning Paths in Physics and Mathematics: Models, Experimental checks, Statistics.* The course was active in Siena from 2005/2006 to 2007/2008 and was attended by thirty teachers.

**Updating course for professional development:** In the last year of the project, we preferred to realize an updating course for teachers in Arezzo at the local high school titled *The nature of light: from classical physics to quantum physics.* In this course, 17 teachers were enrolled and more than an half of scheduled time was spent in physics laboratory.

**Orienting laboratory for classrooms:** Didactic laboratories of the department became available for entire classes to perform some experiments in mechanics, optics and electromagnetism, after a joint programming activity with teachers, to support schools where there are no laboratories or where physical experiences can be performed by teachers for demonstrations but not by students for lack of space and resources. The activity took place at the department for schools in the Siena sorrounding, while for the others, activities at the school were performed by using the equipment borrowed from the department.

**Laboratory of excellence:** The laboratory is designed for developing and strengthening the interest and curiosity in the physics through paths in physics lab (construction and/or characterization of an experimental apparatus), not suitable for teaching in the classroom. Teachers identified among their students a group of suitable work and proposed or chose between our proposed project, that includes realisation and, if possible, the design of an experiment in elementary physics. Small groups of students especially motivated worked in an experimental project coordinated by the teacher out of usual time for school. The activity took place over a period of several months and can also be concluded with the presentation of results in the classroom, or on the day of orientation of the department of physics.

**Pigelleto's summer school of Physics:** The department took the opportunity of using the accommodation in the natural reserve of Pigelleto in Piancastagnaio on Mount Amiata, thanks to financial support from the Provincial Administration of Siena and the contribution of the Fondazione Monte dei Paschi, to host a select group of students in third and fourth classes. The short training period in the beginning of September included intensive work with the purpose of orientation to physics. Laboratories, problem solving, expert seminars with the aim above all of intriguing, ask questions, be fascinated by the discovery of nature and its laws.

**Under the starry sky:** This laboratory is addressed to students and teachers of secondary schools of the first and second cycle. The primary goal is to allow children to have direct knowledge of the scientific method and its fundamental aspects, through experiences in practical astronomy: observation, collection of data, its processing, the formulation of hypothesis on mathematical models, the prediction of observable effects and their verification. All that is achieved by activities performed by night at the astronomical observatory of the department.

**Physics Olympiads:** A preparatory course for preparing interested students to physics olympics, scholastic competition, was held in two schools involved in the project. The department hosted the provincial selection of physics olimpiads.

**Conferences and exhibitions:** Conferences on physics and surroundings realized in local schools to arouse curiosity and attract attention to the physics and PLS. They were often arranged so that many classes can be present. Exibitions on physics topics were organized in the area and young physicists guided students in descovering the activities.

**Orientation day in the Department of Physics:** Teachers were invited to bring their class or part of them at the openday for orientation. A full day in the department, starting with some talk on research and the profession of physicist as academic or freelance scientist in industry. In the afternoon, students were guided in a visit in didactic and research laboratories.

**National Plan for Science Degree**
In the passage from project to plan many activities have been maintained, particularly those which were already organized as PLS laboratories. The request of the Ministry was that, in the first two years of the plan, every local headquarters would organize at least a laboratory of this kind.

The performed PLS laboratories grouped by type are shown in Table 2. More information can be found, such as where they were held and the origins of the students if different from the place of execution, responsible and main contributors in their design and implementation. It is important to underline that there was a complete sharing of responsibilities and working with teachers. The schools inserted in the plan during these two years have been 18.

PLS laboratories can be for entire class or for small groups of students, curricular or extracurricular, realized in school timetable or not. Every year, several hundreds of students and many tenths of teachers are involved in one or more of such laboratories.

The most successful laboratory is the Pigelleto's summer school of Physics, described in details in the next sub-subheadings. In 2011, the Ministry asked to every Regional Scholastic Office to select the best practice in their region for each disciplinary area of the plan. In Tuscany, the best practice for Physics is Pigelleto's summer school.

| PLS LABORATORIES | Responsible | Faculty members and expert teachers involved in designing and implementation |
|---|---|---|
| **Trying and trying again: laboratories in classroom** | curricular | |
| Measures of natural radioactivity | Quattrini* | Montalbano |
| A learning path on heat in vocational schools | Montalbano | De Nicola* Di Renzone* Frati* |
| Physics in winery | Benedetti* | Mariotti Montalbano |
| **Laboratories for approaching the discipline and developing vocations** | extracurricular | |
| Under the starry sky Siena, Grosseto, Arezzo | Millucci | Marchini |
| Toys and physics Arezzo | Porri* | Mariotti |
| Energy from wind Siena | Montalbano | Valentini* Di Renzone* |
| **Deepening laboratory for motivated and talented students** | extracurricular | |
| Detection and measurement of light curves of astronomical images Siena (from Grosseto) | Marchini | Porri* Millucci |
| Waves and energy Siena (from Grosseto) | Montalbano | Di Renzone* Frati* |
| A learning path on spectroscopy Siena (from Grosseto) | Mariotti | Di Renzone* |
| Sound and surrounding Siena (from Grosseto) | Montalbano | Di Renzone* Frati* |
| **Pigelleto's Summer School of Physics** selected like first best practice 2011 for PLS laboratories in Tuscany for physics area (from all provinces) | Organizing commetee | extracurricular |
| | Benedetti* Mariotti Montalbano Porri* | Gargani* Quattrini* 5-10 teachers active in lab |

*Table 2: PLS laboratories implemented in National Plan for Physics in southern Tuscany are presented grouped for type, where stars indicate teachers.*

Some interdisciplinary activities have been realized and a brief description for each one is given in table 3. The most significant interdisciplinary experience, in my opinion, has been the laboratory on modeling, whose legacy has been collected from a up-dating course for teachers in which the issues were focused on learning paths in which the teaching of mathematics and physics are strongly correlated and adapted to the demands of the recent reform of Italian high school.

| Interdisciplinary Activities | Responsible |
|---|---|
| **Orienting laboratory of Science Faculty: education and research** | A. Donati PLS Chemistry |
| Two-days full immersion stage for orienting students in science degrees (collaboration with Chemistry and Mathematics Plans) | |
| **Laboratory of Modeling** | E. Mariotti PLS Physics |
| Workshop for teachers of physics and mathematics on modeling, clarifying topics common to mathematics and physics (collaboration with Mathematics Plans) | |
| **Master in Educational Innovation in Physics and Orienting, University of Udine** | M. Michelini (national) PLS Physics Udine V. Montalbano (local) PLS Physics Siena |
| Inter-university master for teachers in which 19 Universities collaborate for giving courses in laboratory, often focused on a laboratory performed in the National Plan for Science Degree | |
| **Teaching Mathematics and Physics in reformed secondary school** | M. A. Mariotti PLS Mathematics |
| Updating Course for Teachers at School on adapting mathematics and physics teaching in interdisciplinary designed learning paths for reformed vocational and high schools (collaboration with Mathematics Plans) | |

*Table 3: Interdisciplinary Activities implemented in National Plan for Physics in southern Tuscany.*

**The Pigelleto's Summer School of Physics:** The summer school has reached the sixth edition and appears to be the activity more interesting and enjoyable for students both for form of organization that for contents.

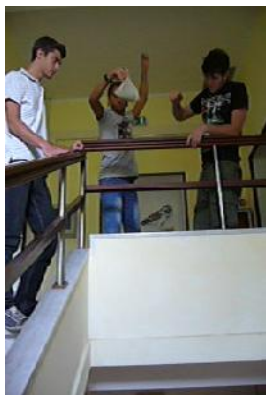

*Figure 5: Students perform an experiment on energy transformation at Pigelleto's summer school.*

Forty students from high school are selected to attend at full immersion summer school of physics in the Pigelleto Natural Reserve, on the south east side of Mount Amiata in the province of Siena. The school begins usually in early of September and last for four days. The 2011 edition was titled *Thousand and one energy: from sun to Fukushima,* some previous editions were *Light, color, sky: how and why we see the world* ((2006), *Store, convert, save, transfer, measure energy, and more…* (2007), *The achievements of modern physics* (2009), *Exploring the physics of materials* (2010). Topics are chosen so that students are involved in activities rarely pursued in high school, relationship with society are outlined and discussed. The students are selected by their teachers in the network of schools involved in the National Plan for Science Degree [8].

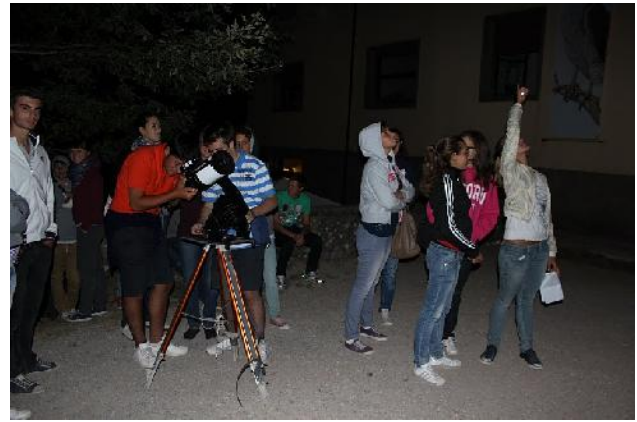

*Figure 6: In the evening, the stars look down…and student are watching them.*

In the morning lessons are proposed in which the necessary background for the following activities in laboratory is given. In the afternoon small groups of students from different schools and classes are engaged in laboratories where are forced to take an active role. All groups are supported by one or two teachers that are available to discuss any idea.. Usually we propose different laboratories for each group and the task of preparing a brief presentation for other students is given in order to share with them what they have learned. After dinner, an evening of astronomical observation of the sky is usually expected. If it is cloudy, a problem solving evening is proposed.

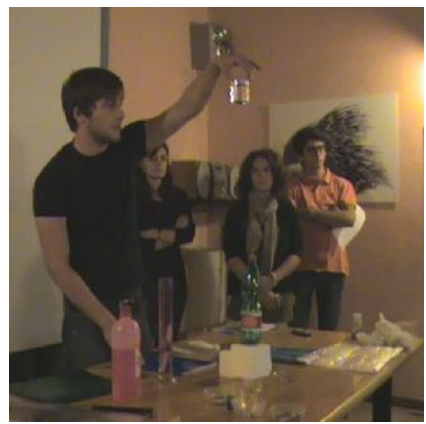

*Figure 7: A group of students present an experience on mechanical and optical properties of growing spheres*

In designing summer school activities, we pay attention to several aspects that can render more effective this action both for teachers that for students. Let me give some example :
- the main topic is related to all activities and must be not trivial;
- when it is possible, laboratories are made with poor materials or educational devices provided by some schools in such way that teachers can duplicate easily the lab in their school;

- almost all laboratories lead to at least one measure and its error valuation;
- methodologies are discussed and selected with the teachers involved in Physics Plan;
- in order to have the best collaboration, students' groups are inhomogeneous and formed by following the teachers suggestions;
- usually in laboratory an expert and a young teacher are involved in order to improve teacher practise.

The summer school is a full interdisciplinary action because several different skills of students are engaged, such as physics and mathematical ones, social behavior for collaborate actively and efficiently with other students, communicative ones for a good transmission of information between groups and in presentation for sharing knowledge. Furthermore, very often the main topic is interdisciplinary from the very beginning, such as materials science, or becomes so in discussing issues relevant to society, for example in the case of energy sources or nuclear energy.

*Modeling* Since in 2009 also the Department of Mathematical Sciences of University of Siena launches its plan, a workshop for teachers of physics and mathematics on modeling was performed in collaboration with Physics Plan.

This activity is inter-disciplinary for constructions and has continued in an updating course for teachers in which selected topics, named in the same way in both disciplines, are discussed in order to design interdisciplinary learning paths. The purpose is to clarify these topics by using specific tools from physics and mathematics and to outline the similarities and the differences in both contexts.

We believe that this activity can be useful for students, which can acquire a more profound insight on some fundamental concepts, and for teacher professional development.

### 4. CONCLUSIONS

The actions realized in these years in order to maintaining and increasing the number of motivated and talented students who enroll in courses scientific degrees in Italy are beginning to give some result, as shown in figure 8.

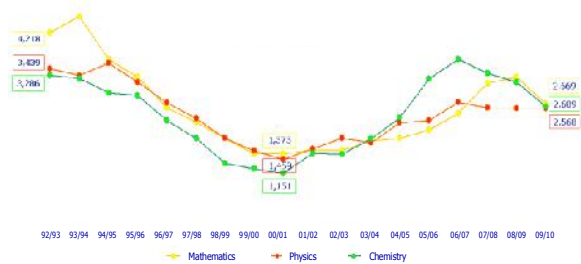

*Figure 8: Enrollment in Italian universities in basic sciences[7]*

Furthermore, many teachers are interested in activities in which new interdisciplinary methodologies are developed for increasing students' understanding in high school of selected topics in which both Mathematics and Physics give a contribution in the construction of important and mutually reinforcing basic concepts of the two disciplines.

A goal achieved by the National Plan for Science Degree is that a network of schools and teachers has established and it is permanently active in improving the teaching of sciences in high school. A further result is that many teachers, even if forced by the recent reform of the secondary school, are beginning to collaborate actively in mathematics and physics education research.

### 5. ACKNOLEDGEMENTS

This work is based on activities and experiences which were performed within the National Plan for Science Degree supported by Italian Ministry of Education, University and Research. The author would like to thank the director of department of physics, Angelo Scribano, for the support.